\documentclass[12pt,letterpaper]{article} 
\usepackage[margin=1in]{geometry}
\usepackage{setspace}

\usepackage{amsmath} 
\usepackage{titlesec}
\usepackage{graphicx} 
\usepackage{caption}
\usepackage{enumitem} 
\usepackage{natbib} 
\usepackage{hyperref} 
\hypersetup{
    colorlinks=true,   
    linkcolor=blue,   
    citecolor=blue    
}
\usepackage{fancyhdr}  
\fancypagestyle{plain}{
    \fancyhf{}  
    \fancyfoot[R]{\footnotesize \thepage}

}

\title{\textbf{\large Exploring Seismic Signal Detection and Source Identification of Atmospheric Entries: The Hayabusa2 Sample Return Capsule as a Benchmark}}

\author{\normalsize Iona Clemente\textsuperscript{1,2}, Eleanor K. Sansom\textsuperscript{2}, Hadrien A. R. Devillepoix\textsuperscript{1,2},\\ \normalsize Taichi Kawamura\textsuperscript{3}, Benjamin A. Fernando\textsuperscript{4}, Raphael F. Garcia\textsuperscript{5}, Olivia Collet\textsuperscript{6}
}

\date{}

\begin{document}
\maketitle
\thispagestyle{plain}

\small
\noindent \textsuperscript{1} Space Science and Technology Centre, School of Earth and Planetary Science, Curtin University, Perth, Australia.
\noindent \textsuperscript{2} International Centre for Radio Astronomy Research, Curtin University, Perth, Western Australia, Australia.
\noindent \textsuperscript{3} Institut de Physique du Globe de Paris, 35 Rue H\'el\`ene Brion, Paris 75013, France.
\noindent \textsuperscript{4} Department of Earth and Planetary Sciences, Johns Hopkins University, Baltimore, Maryland, United States.
\noindent \textsuperscript{5} Institut Sup\'erieur de l’A\'eronautique et de l’Espace (ISAE-SUPAERO), Universit\'e de Toulouse, Toulouse, France.
\noindent \textsuperscript{6} Centre for Exploration Geophysics, Curtin University, GPO Box U1987, Perth, WA 6845, Australia.\\

\noindent Corresponding author: iona.clemente@postgrad.curtin.edu.au\\

\section*{Abstract}
\hspace{5mm}  This exploratory study investigates whether seismic signals can be used to infer fragmentation during a fireball event. Re-entry objects, particularly sample return capsules (SRCs) such as the one from the Hayabusa2 mission, behave similarly to slow meteors during atmospheric entry and provide valuable insights into natural fireball events. In this study, we initially analyse seismic signals from the Hayabusa2 SRC re-entry, which took place on December 5, 2020, over South Australia. The SRC’s signature was captured by a dense network of seismic stations \citep{eakin2018lake, o2020ausarray}, offering a unique opportunity to investigate the signals’ characteristics and verify their connection to the re-entry event. The ballistic trajectory  was confirmed as the source shock mechanism for this event. We isolate this signal and use it as a reference for a ballistic shock signature and compare it to three other fireball case studies, including a suborbital re-entry and two natural meteoroids. Although factors such as local geology and atmospheric conditions were not considered in this preliminary study, our results show promise, with high correlations for events with purely ballistic trajectories and lower correlations for those involving fragmentation or airbursts. This implies that seismic data may be able to disambiguate whether any particular fireball event underwent significant fragmentation or airburst, key phenomena for assessing body strengths.

\section*{Introduction}
\hspace*{5mm} When meteoroids pass through the Earth’s atmosphere, they rapidly decelerate and can produce bright fireball phenomena. In most cases they break up and entirely vaporise \citep{ceplecha2005fragmentation}, though some may survive longer and even drop meteorites on the ground. If the objects are large enough (greater than 10 centimetres diameter), they can generate shock waves \citep{bronshten1964problems, revelle1976meteor, revelle2008acoustic, silber2018physics} that can be recorded by ground-based instruments such as seismic and infrasound sensors.

\noindent Shock waves can be produced by different mechanisms that can occur at various points along a meteoroid’s trajectory. (1) The first mechanism is the \textit{hypersonic trajectory} of the meteoroid through the atmosphere itself, which creates a conical bow shock known as a Mach cone \citep{revelle1976meteor}. As meteoroids enter Earth’s atmosphere at velocities ranging in the order of 11-73 km/s (equivalent to Mach number 35 to 240: \citet{ceplecha1998meteor, murad2002meteors}), their velocities are significantly higher than the speed of sound. Consequently, the Mach cone angle is sufficiently small to allow the cone’s wavefront to be approximated as a cylinder \citep{tsikulin1970shock, revelle1976meteor}. Such a shock wave’s propagation is assumed to be perpendicular to a meteoroid’s trajectory, and will mean that a direct arrival from the ballistic trajectory will match the time of arrival to the shortest distance to the trajectory. (2) Shock waves are also generated when the force exerted by the motion of the meteoroid through the atmosphere (ram pressure) on its surface exceeds its compression strength \citep{cevolani1994explosion, stevanovic2017bolide}, causing it to break into smaller fragments. This phenomenon is called a fragmentation event \citep{ceplecha1998meteor, revelle1976meteor, silber2018infrasound}. (3) The last phenomenon that can cause shock waves is a catastrophic final airburst, which corresponds to the complete disintegration of the meteoroid into fine dust \citep{klekociuk2005meteoritic}. Both fragmentation events and final airbursts produce shock waves that propagate quasi-spherically \citep{edwards2008seismic}. These latter two mechanisms can be indicative of object strength, and are not expected to be generated by human-made sample return capsules (SRCs) as they are specifically designed to withstand their passage through the atmosphere and reach the ground intact. 

\noindent In some cases, typically those of higher energy, the shock wave can couple with the ground and produce seismic surface and body waves \citep{revelle2004entry, edwards2008seismic}. In addition to recording direct acoustic waves, seismic sensors are also capable of detecting these air-to-ground coupled waves. Usually, seismic expression is limited to surface waves, such as Rayleigh waves. Only the most energetic events generate body waves (P- or S-waves) that can be observed above the noise level. Examples include the Chelyabinsk event \citep{tauzin2013seismoacoustic, brown2003moravka, stevanovic2017bolide, karakostas2015inversion}, or rarely documented cases of direct impacts \citep{brown2008analysis, pichon2008evidence, tancredi2009meteorite}.\\

Around 250 fireballs enter Earth’s atmosphere every day, given the decade of fireball data from the Australia’s Desert Fireball Network (averaging one event per night with a coverage area of ~0.4\% of the Earth; \citet{devillepoix2022trajectory}).  However, their occurrence is sporadic —fireball events are unpredictable in both time and location. Passive, large-area surveillance networks are required to monitor these events effectively. The Desert Fireball Network (DFN), has had nearly 10 years of observing Australian skies from up to 52 observatories, covering an area of 3 million km$^2$. Its primary goal is to detect fireballs and calculate their trajectories and orbits to recover meteorites \citep{devillepoix2019observation, devillepoix2018dingle}. Since its installation, the DFN has triangulated more than 2,000 events across Australia, including the return of the Hayabusa2 SRC. 

\noindent In contrast to meteors and fireballs, the re-entry of sample return capsules is a much rarer event. Since the first SRC re-entry of NASA’s Genesis in September 2004 \citep{revelle2005genesis}, there have only been four other re-entries: Stardust \citep{revelle2007stardust}, Hayabusa \citep{ishihara2012infrasound}, Hayabusa2 \citep{nishikawa2022modeling, sansom2022scientific}, and most recently OSIRIS-REx in September 2023 \citep{silber2025along, fernando2024seismoacoustic}. The last four events were monitored using seismic sensors.

\noindent Because trajectories of re-entry events are less energetic than natural fireballs, capturing their shock wave signals can be more challenging. However, studying these events remains crucial as SRCs can be considered artificial meteors, entering the Earth’s atmosphere at the lower limit of the velocity range for meteors (12 km/s). They allow instruments to be deployed at a known time and location, and at a scale and focus that would not be feasible for sporadic fireballs. They also allow us to gain insights and deepen our understanding for natural meteoroid entries.  \\

As SRCs are not expected to undergo fragmentation or airburst, direct airwave signals are hypothesised to originate from the ballistic trajectory (Shock Mechanism 1; SM1). 
We further hypothesise that seismic signatures recorded from other fireball events where SM1 is the dominant source may exhibit common characteristics. This would mean that fireball events lacking energetic fragmentation or airbursts may be distinguishable from cases where such phenomena do occur (potentially involving SM2 or SM3). This would allow seismic data the capability of ball-parking whether any particular fireball event underwent significant fragmentation/airburst, key phenomena for assessing body strengths. \\

In this study we test the initial hypothesis by analysing seismic records from the re-entry of the Hayabusa2 SRC, using data from the Geological Survey of South Australia’s temporary 5G/6K AusArray networks \citep{eakin2018lake, o2020ausarray} which were operational at the time. The suite of analyses we perform aims to examine the source of the shock and its connection to the object’s trajectory, supported by a visualisation tool based on the filterbanks technique. 

\noindent Following this, we then compare the Hayabusa2 signal to those from other case studies. Cross-correlation techniques are commonplace for analysing signals from a single event across distributed sensors. Here we apply this method to instead compare different events, with the aim of probing signal similarities that may imply similar shock mechanisms. This is an exploratory study based on four case study events, intended to investigate whether seismic signals could be used to disambiguate the source shock mechanism and infer whether a fireball likely underwent fragmentation. This is particularly valuable for events where no optical data were recorded, or there is limited seismic sensor coverage. As most fireballs are from unplanned, sporadic events, the density of sensors will greatly vary. The full suite of analyses conducted here for the Hayabusa2 case is rarely feasible. Being able to extract such information as likely shock mechanism from a single seismic sensor would be extremely valuable. \\

\section*{Case study events}
\vspace*{3mm} 

\noindent \textit{SRC re-entry: Hayabusa2}

On December 3, 2014, the Hayabusa2 mission was launched toward the asteroid 162173 Ryugu. To study this near-Earth asteroid, Hayabusa2 was tasked with collecting samples from its surface and bringing them back to Earth \citep{fujita2011overview}. Nearly six years later, on December 5, 2020, the sample return capsule of the Japanese mission returned to Earth, with a planned landing in the Woomera Prohibited Area in South Australia. To record this predicted re-entry, various sensors were deployed, including cameras, radio antennas, seismic and infrasound sensors \citep{sansom2022scientific, nishikawa2022modeling}. The SRC re-entered the atmosphere at approximately a 12° angle. It started its bright flight (luminous phase) at 73 km altitude at an entry velocity of 11 km/s. The bright flight lasted approximately 26.5 seconds, and covered 232 km, ending at an altitude of 41 km with a speed of 5 km/s.\\

\noindent \textit{Natural event: Fireball DN210112\_02}

The DN210112\_02 fireball event took place on January 12, 2021, over Lake Torrens, South Australia, and was captured by eight Desert Fireball Network (DFN) cameras \citep{howie2017build, devillepoix2020global}. Among the more than 2,000 fireball events observed and triangulated by the DFN, this event was specifically selected because it also entered over the temporary 5G/6K AusArray networks \citep{eakin2018lake, o2020ausarray} during their operational period. Data reduction and triangulation were performed using established methods within the DFN pipeline (e.g. see \citet{devillepoix2022trajectory} and references therein). As expected for a natural event, its initial speed was significantly higher than for the SRC re-entries, entering the atmosphere at 37 km/s. Bright flight started at 95 km altitude, ending at 48 km. The bright flight lasted only 3 seconds, with a slope of 27 degrees, ending with a speed of 22 km/s. This event is unlikely to have any surviving material (no meteorites). Although its orbit suggests a potential cometary origin (Tisserand parameter of 1.6), it did not display any significant fragmentation or airburst events. This event will allow us to compare our method for ballistic signal identification with a natural event.\\ 

\noindent \textit{Human-made space debris: Soyuz re-entry}

On August 7, 2023, multiple witnesses in Melbourne and Tasmania (Australia) reported observing a fireball and hearing a sonic boom, which were traced back to the re-entry of the upper stage of a 2.5-ton Russian Soyuz 2.1b rocket. The rocket had originally delivered a GLONASS navigation satellite into orbit. The debris was intended to be discarded into the Southern Ocean, and a low chance of fragmentation was expected.

\noindent In addition to capturing natural fireball events, such as DN210112\_02, DFN cameras can also record the re-entry of human-made objects into Earth’s atmosphere. This capability was demonstrated in December 2020 for the Hayabusa2 SRC re-entry, when 17 DFN cameras were used to triangulate its trajectory \citep{sansom2022scientific}, as well as capturing the beginning of the Soyuz re-entry. The Soyuz, being a sub-orbital launch vehicle, entered at a lower speed of 7.3 km/s and slowed to a final velocity of 6 km/s, with a trajectory slope of 1.7 degrees. The bright flight began at an altitude of 78 km.\\

\noindent \textit{Natural event: Queensland fireball}

On May 20, 2023, a large meteoroid entered Earth’s atmosphere over North Queensland, Australia. This fireball was detected from orbit by the United States government sensors (Center for Near Earth Object Studies (CNEOS), \textit{cneos.jpl.nasa.gov}), which estimated its energy release at 7.2 kilotons of TNT, making it the 5th most energetic event over land in the world. Casual video and surveillance footage of this event were used to re-create the trajectory, which has been validated by Silber et al. (in prep). There was a significant final airburst, with likely other fragmentation events throughout the trajectory. We would expect signals from this event to differ from all other cases.

\section*{Data and Methods}
\begin{figure}[h!]  
    \centering
    \includegraphics[width=0.98\textwidth]{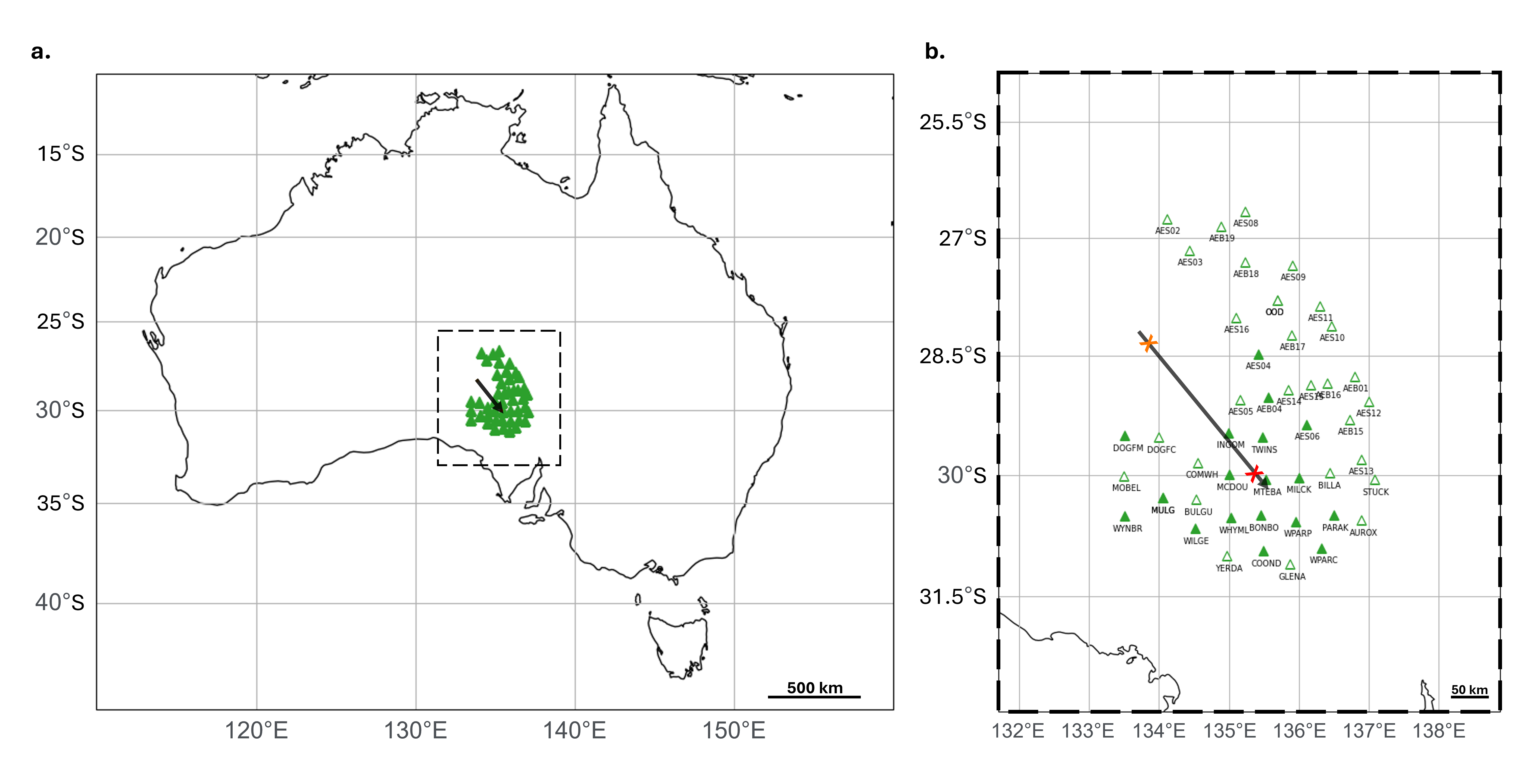}
    \caption{Hayabusa2 sample return capsule re-entry trajectory. \textbf{a.} Large-scale view of the re-entry trajectory over a dense network of seismic stations (green triangles). \textbf{b.} Zoomed-in view: X symbols mark the points along the SRC trajectory, with orange indicating the start of the bright flight (luminous phase during atmospheric entry) and red marking the end. Filled green triangles represent the 18 stations selected for the Hayabusa2 re-entry study.}
    \label{fig1}
\end{figure}

\hspace{5mm} The main data sources used are  seismic data and triangulated fireball trajectory data, for each of the case studies. Raw seismic data were acquired for all the case study events by addressing a request through the webservice provided by FDSN \citep{fdsn} to the Australian National Seismograph Network (ANSN) to access the ‘AU’ network \citep{geoscience_australia}. Stations in the AU network are sparsely and unevenly distributed, with only a few stations located close enough to detect direct-coupled airwaves for most of the studied events (see Table 1). In addition to these, by acquiring permission from data holders, the access to additional data from two restricted networks, ‘5G’ \citep{eakin2018lake} and ‘6K’ \citep{o2020ausarray}, was obtained. These networks are high-density seismic networks with limited operational time, running from 2018 to 2022 and 2020 to 2022, respectively. Their coverage includes the Hayabusa2 SRC re-entry case, as well as the natural DN210112\_02 event. The data from the 5G and 6K networks are a unique opportunity to investigate the airwave signals across a dense array of sensors (Figure \ref{fig1}). The Python language \citep{rossum2009python}, and its dedicated seismic library ObsPy \citep{beyreuther2010obspy, krischer2015obspy} were used to process and analyse these seismic data. This involved using the obspy.signal module to remove the instrument response, convert data to velocity, and apply filtering as detailed below. The infraGA package\footnote{\url{https://github.com/lanl-seismoacoustics/infraga} (last viewed 14 february 2024)} \citep{blom2012impulse, blom2017modeling} was used to confirm that all recorded sources were from direct airwaves, with no identifiable waveguides reaching the ground, using the atmospheric model for the Hayabusa2 re-entry region of \citet{nishikawa2022modeling}. The trajectory information of the Hayabusa2 re-entry and of DN210112\_02 used in this study were provided by the Desert Fireball Network.\\

Initially, we wish to conduct an in-depth analysis of the Hayabusa2 SRC re-entry where the density of seismometers allows for a full suite of analyses. This will validate our signal identification method for other events, and validate the source shock mechanism from the ballistic trajectory. We began by utilising a visualisation tool based on the filterbanks technique which allows us to identify the frequency range where signals are clearly defined above noise. Subsequently, we proved the source origin to the trajectory of the SRC re-entry using established methods such as wave velocity estimation, cross-correlations and polarisation analyses.\\

Following on from a successful confirmation of SM1 for the Hayabusa2 case, we will use the isolated SM1 signal and use it as a reference for other events. 
In order to perform our comparison, we first identify event-related signals for other case studies using the same filterbanks technique used in the Hayabusa2 case. 
To then compare events, we do a cross-correlation between the Hayabusa2 SRC re-entry with available data from each case study to investigate similarities in signals and potentially constrain the source shock mechanism.\\

\noindent \textbf{Filterbanks technique}
\vspace*{3mm} 

\begin{figure}[h!]
    \centering
    \includegraphics[width=0.98\textwidth]{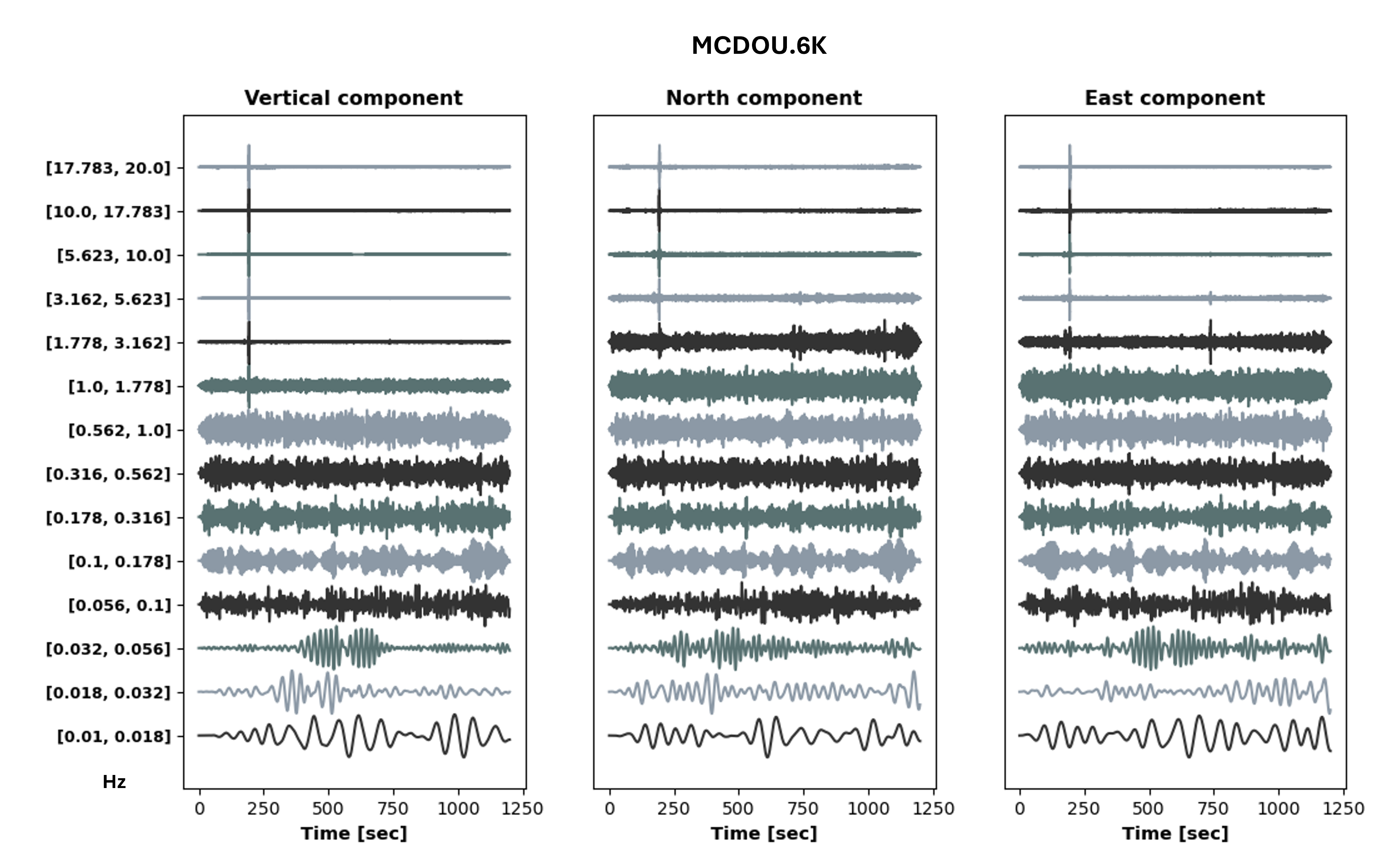}
    \caption{Application of the filterbanks method on seismic records of the Hayabusa2 re-entry from the MCDOU (6K) station, a 3-component seismometer (Vertical, North, East). Fourteen bandpass filters have been applied, with equal increments in log space. These in intervals of \( 0.01 \times 10^{0.25n} \) where \( n \) increases by 1 for each bandpass. The x-axis corresponds to the elapsed time since the start of the event (17:28:54 UTC). Note, the frequencies above 1 Hz show the greatest signal-to-noise ratio, and will be used to define the lower limit of the bandpass filter for this event.}
    \label{fig2}
\end{figure}

This method provides an approach for visualising seismic data across different frequencies. It draws inspiration from the work of \citet{butterworth1930theory, duval2000filter}, as well as the filterbanks technique used by the InSight team to plot Marsquake signals at individual frequencies \citep{clinton2021marsquake, ceylan2022marsquake, kawamura2023s1222a}. In our approach, after removing the instrumental response, the data are filtered into narrow frequency bands ranging from 0.01 to 20 Hz using the ObsPy bandpass filter, implemented as a one-pass, four-pole Butterworth filter with a roll-off rate of 24 dB per octave beyond the -3 dB corner frequencies (\citet{beyreuther2010obspy}; e.g. Figure \ref{fig2}). We selected this frequency range because it allows us to capture both long-period seismic waves, such as surface waves, and high-frequency waves such as direct airwaves, enabling the detection of all types of waves generated by an atmospheric re-entry. To ensure evenly spaced frequency bands on a logarithmic scale, the frequencies are represented as power of 10, with an exponent value increasing by 0.25 at each step, starting from 10$^{-2}$ (0.01 Hz) up to the upper limit of 20 Hz.\\

\noindent This method also allows us to visually identify frequency bands with detectable signals and a favourable signal-to-noise ratio for analysis. When multiple stations are within range of a fireball trajectory, the station with the lowest overall noise is used as a reference station to determine the optimal frequency bounds of the bandpass filter to study a given event.
For example, in the Hayabusa2 case, a bandpass filter with a range of 1 to 19 Hz is chosen based on the MCDOU filterbanks visualisation (Figure \ref{fig2}) and applied to filter data from all other stations within a 300 km radius around the midpoint of the SRC re-entry trajectory. For signal comparisons in this study, data from the vertical component are used, and the 1-19 Hz bandpass filter is applied to all stations being compared.\\

\begin{figure}[h]
    \centering
    \includegraphics[width=0.98\textwidth]{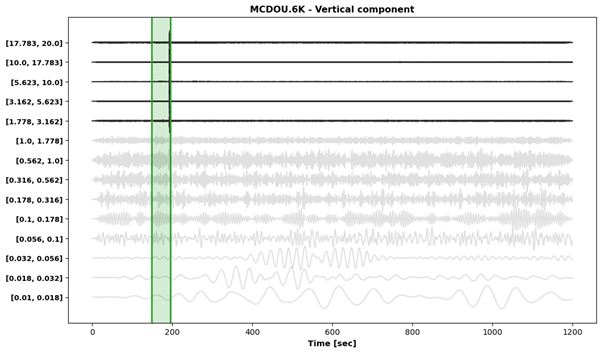}
    \caption{Hayabusa2 SRC re-entry expected arrival time at MCDOU (6K) seismic station, with reference to the filterbanks filtering of the vertical signal component. The green rectangle represents the time window during which a direct shock wave would be expected, given the shortest distance to the station, and a minimum (263 m/s) and a maximum (347 m/s) sound wave speed. The higher frequencies show the higher SNR. The x-axis corresponds to the elapsed time since the start of the event (17:28:54 UTC).}
    \label{fig3}
\end{figure}

To constrain an approximate time window for the expected arrival of the acoustic wave to each station, we first calculate the shortest distance between the seismic station and the bright flight of the SRC, which represents the luminous phase during atmospheric re-entry. The latest arrival (right boundary in Figure \ref{fig3}) is calculated using this distance and a sound speed of 263 m/s, corresponding to the sound speed at an altitude of 72.9 km (start of the SRC luminous flight), based on the atmospheric model for the Hayabusa2 re-entry region \citep{nishikawa2022modeling}. For the earliest arrival (left boundary in Figure 3), the shortest distance is again used, but in combination with the maximum sound speed of 347 m/s, representing the sound velocity at ground level from the same atmospheric model. At this stage, a direct arrival, and therefore, straight-line propagation is assumed.\\

\noindent \textbf{Validating the signal from filterbanks: Hayabusa2 case}
\vspace*{3mm} 

\begin{figure}[h]
    \centering
    \includegraphics[width=0.73\textwidth]{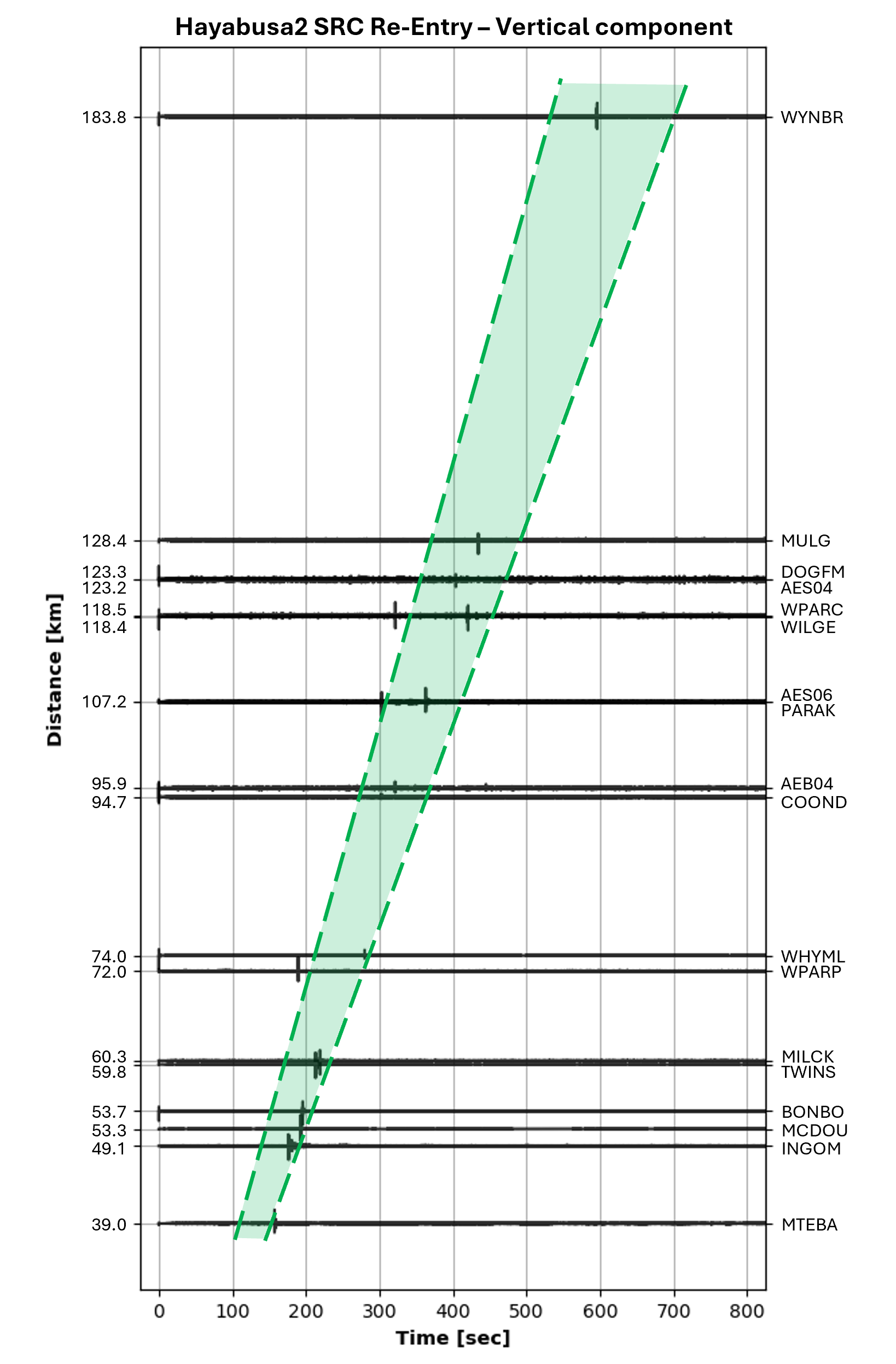}
    \caption{Hodochrone of seismic signals observed across the selected stations during Hayabusa2 SRC re-entry, plotted relatively to their shortest distances from the event (y-axis). The seismic traces are bandpass filtered between 1-19 Hz and represent the normalized vertical velocity. The green window marks the time frame in which the shock waves are expected to arrive at the stations, with a lower boundary indicating the highest possible sound speed (347 m/s) and the upper boundary indicating the lowest possible sound speed (263 m/s). The x-axis shows the elapsed time since the start of the event (17:28:54 UTC). \textit{Note:} Some stations have very similar distances to the re-entry trajectory, causing their traces to overlap on the figure, though they represent two distinct signals.}
    \label{fig4}
\end{figure}

Eighteen seismic stations with clear signals and high signal-to-noise ratios (SNR) were selected to study the Hayabusa2 SRC re-entry. Since signals have been found within the time window of the shock wave’s arrival (cf. Figure \ref{fig4} they are considered potential candidates for originating from the Hayabusa2 re-entry. To confirm this source and categorize the signals, three validation steps were defined, as outlined below:\\

\noindent \textbf{\textit{1. Average apparent wave velocity computation}}
\vspace*{3mm} 

Estimating the apparent velocity of the wave recorded at the seismic stations helps us identifying the type of wave being observed. The aim here is not to determine an exact velocity, but rather to assess whether the wave is likely to originate from an atmospheric, acoustic source or a ground-coupled wave. Effects such as wind, atmospheric temperature and pressure variations will affect the former, while local geology will affect the latter. However, an atmospheric source would travel at velocities closer to the typical speed of sound ($\approx 300$ m/s), whereas seismic waves are generally an order of magnitude faster, making them distinguishable. To calculate the apparent wave velocity at each of the 18 selected stations, we used the following well-known relation:

\begin{equation}
v = \frac{d}{t}
\end{equation}

\noindent where \textit{d} is the shortest distance between the station and the re-entry trajectory, and \textit{t} is the wave’s arrival time (in seconds elapsed since the start of the event at 17:28:54 UTC).\\
\noindent We then computed the average of the wave velocities obtained across all stations. This approach provides an estimate of the apparent wave’s velocity, allowing us to assess whether it is on the order of the sound speed or closer to typical seismic waves velocities. If the velocity approaches the speed of sound, we employ shock propagation modelling using infraGA \citep{blom2024} to investigate whether the signals are direct arrivals or if they suggest the presence of an atmospheric waveguide.\\

\noindent \textbf{\textit{2. Cross-correlation analysis}}
\vspace*{3mm} 

To confirm the source of the signals for the Hayabusa2 SRC, we take advantage of the high-density network to assess the similarities between signals from stations located at different distances from the re-entry event. A cross-correlation analysis was performed by sliding one over another and calculating the product of their values at each time shift \citep{shapiro2004emergence, campillo2008long, curtis2006seismic}, followed by normalisation of the result. This analysis helps us verify if nearby stations are detecting the same event and, most importantly, if all detected signals can be confidently attributed to the Hayabusa2 SRC re-entry.\\
\noindent To do so, for each pair of stations, one signal is selected and serves as the reference, with a focus on a 6-second window centered around its peak absolute amplitude. The second station’s signal spans 500 seconds. The correlation between the two signals is performed by sliding the 6-second reference signal across the full length of the 500-second signal. The resulting correlation coefficients (ranging from 0 to 1 (or -1)) quantify the similarity between each combination of signals. These coefficients are compiled into a histogram for further analysis.\\

To assess the significance of the correlation coefficient obtained when the reference signal correlates with the 6-second window of the second signal containing the presumed shock wave arrival, we must account for the contribution of background noise. To do this, we employed the time-reversed template approach developed by \citet{slinkard2014multistation}. Since the reference signal is not expected to strongly correlate with the time-reversed signal, this method helps establish a threshold coefficient where the correlation begins to trigger on noise. We computed this correlation coefficient threshold using a false alarm rate (FAR) set at 0.1\%. This means that a correlation coefficient exceeding the threshold has only a 0.1\% chance of being due to noise. In our case, the \citet{slinkard2014multistation} method involves time-reversing the 500-second signal and applying the correlation process in this reversed state. The correlation coefficient from the forward correlation is then compared against the threshold. If the coefficient exceeds the threshold, the correlation is deemed significant, as it surpasses the false alarm rate.\\

Signals recorded by distributed sensors of the same event will be somewhat affected by the local geology as well as distance to the source. We do not expect perfect correlations, but by identifying those that are statistically significant, we hope to distinguish between the main source shock mechanisms.\\

\noindent \textbf{\textit{3. Polarisation analysis}}
\vspace*{3mm} 

Polarisation analysis determines the direction of arrival of a signal relative to the seismic station where it was recorded \citep{bormann2013seismic}, providing valuable insights into its origin. Using the ObsPy’s particle motion function \citep{beyreuther2010obspy}, the motion is visualised as a vector, revealing both the direction and amplitude of the wave’s movement. In this study, we analysed particle motions from selected stations to help identify the origin of the observed signals. It is expected that if the motions for all stations converge toward the same location (sensor location agnostic), it would suggest a point source with a quasi-spherical shock produced by a fragmentation or airburst event. In contrast, for a ballistic shock source mechanism, the shock wave is generated continuously along the trajectory. Particle motions for each sensor will point toward the shortest distance to the trajectory, and therefore appear to come from different locations along the re-entry path. \\

\noindent Particle motions were computed using only the horizontal components of the sensors (North and East), as we are primarily concerned with the azimuth angle.  They were calculated for a focused window around the first and strongest arrival of the signals, with a duration of 0.5 seconds. We found that varying the window size did not significantly affect the results. To ensure clarity in visual representation, particle motions were normalized across all stations in the figures. Polarisation of the signals are also affected by local geology and incidence angle of the source, though the azimuths of particle motion are mostly unaffected.\\

This technique is applied to the dense Hayabusa2 SRC data only. Initially, we considered using polarisation analysis to determine the source of the signals for the other case study events as well. The Hayabusa2 SRC re-entry provided an ideal and relatively unique setup with a dense and well distributed seismic network, which made the polarisation feasible. However, this was not the case for the other events. For instance, even though the DN210112\_02 fireball occurred in the same region as Hayabusa2, it entered at the edge of the dense network, with all stations positioned on one side of the fireball’s trajectory. This asymmetry made interpreting the polarisation results challenging. We encountered similar issues with stations distribution for the Soyuz re-entry and had only one available station for the Queensland fireball. This will likely be the case for the majority of fireball events recorded by seismometers. As a result, beyond using it to validate the results of Hayabusa2 SRC, we decided to not pursue polarisation analysis for the signals sources identification.\\
\noindent Given that the Hayabusa2 SRC signal source was expected to be linked to its ballistic trajectory (and was confirmed in the Results section), we opted to compare signals’ origin from the different events using cross-correlations instead.\\

\noindent \textbf{Cross-event signal correlations}
\vspace*{3mm} 

As part of the validation step described above, cross-correlations were performed between signals from the same event (Hayabusa2 SRC re-entry), recorded by different seismic stations. Here, we present a derived cross-correlation analysis comparing the Hayabusa2 re-entry signal with signals from the other case study events. To achieve this, the signal recorded at MCDOU (6K) station was used as the reference for the Hayabusa2 re-entry. This 6-second reference signal is centred on the peak of absolute maximum amplitude. Signals from the other events were filtered within the same frequency range as the Hayabusa2 re-entry, between 1 and 19 Hz.\\

\noindent We anticipate high correlation coefficients for signals from the ballistic trajectory shock wave. In contrast, signals from disruptions or airbursts may show lower correlation coefficients, depending on how they interact with the ballistic wave at different distances. Here again we evaluate the correlation coefficient significance using the time-reversed template approach \citep{slinkard2014multistation} and the computation of the correlation coefficient threshold value with a false alarm rate of 0.1\%.

\section*{Results}
\hspace{5mm} This section details the results obtained from the analysis of the seismic records of the Hayabusa2 SRC re-entry and provides a comparison between the Hayabusa2 signal with those from the other case study events.\\

\noindent \textbf{Hayabusa2 SRC}
\vspace*{3mm} 

The Hayabusa2 SRC re-entered Earth’s atmosphere on December 5, 2020, at 17:28:54 UTC. Of the 18 stations selected with high signal-to-noise ratios, the first signal of the re-entry was detected by MTEBA (6K) at 17:31:32, approximately 2 minutes 30 seconds after the re-entry began (Figure \ref{fig4}). MTEBA is the closest station to the SRC’s trajectory, with its shortest distance calculated to be 39 km. In contrast, the latest signal was recorded at WYNBR (6K) at 17:38:50, approximately 183.8 km from the trajectory at its closest point. The arrival times of signals detected at all the 18 selected stations, along with their shortest distances to the re-entry trajectory, are presented in Table 2.\\

\begin{figure}[h]
    \centering
    \includegraphics[width=0.98\linewidth]{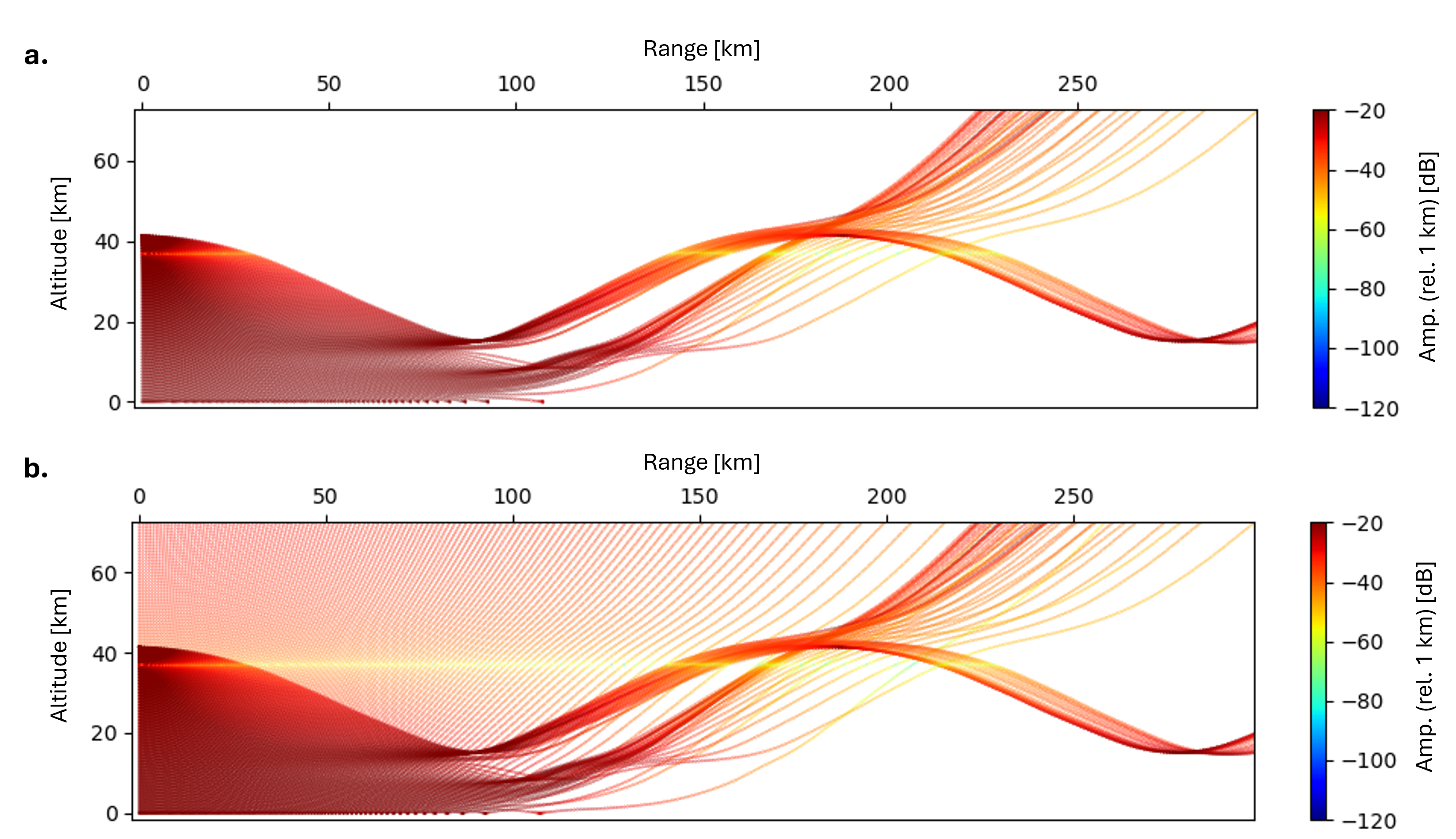}
    \caption{Outputs of infraGA model for ray tracing of shock waves, using the atmospheric model of \citet{nishikawa2022modeling}. The source height of 41 km represents the end of the Hayabusa2 luminous fireball trajectory. Sampling is every 5 degrees from vertical to horizontal, toward an azimuth of 090$^o$ (East). \textbf{a.} Shows direct arrivals only, while \textbf{b.} shows the continued propagation after 'bouncing'. The dark red points at ground level show a single contact point with the Earth.}
    \label{fig5}
\end{figure}

The first step in the validation process was to categorise the type of waves that were recorded. The average wave velocity estimation yielded a velocity of 257 m/s, which is closer to the speed of sound (approximately 340 m/s) than to typical seismic wave velocities, which are on the order of thousands of meters per second. This suggests that the recorded signals likely associated with the atmospheric arrival of the shock wave generated by the SRC re-entry. Additionally, the shock propagation modelling using infraGA \citep{blom2024} also confirmed that all paths to the ground were direct arrivals, reaching up to 150 km from the trajectory. We ran the infraGA model from both the start and end point of the Hayabusa2 SRC luminous fireball trajectory, in all directions. In all cases, the direct arrivals show a single bounce, and any identified waveguides do not reach ground level. Figure \ref{fig5} shows an example output from the end of the trajectory toward the East (azimuth: 090$^o$), showing the direct arrivals only, as well as the path of any bounces.\\

\begin{figure}[h]
    \centering
    \includegraphics[width=0.98\textwidth]{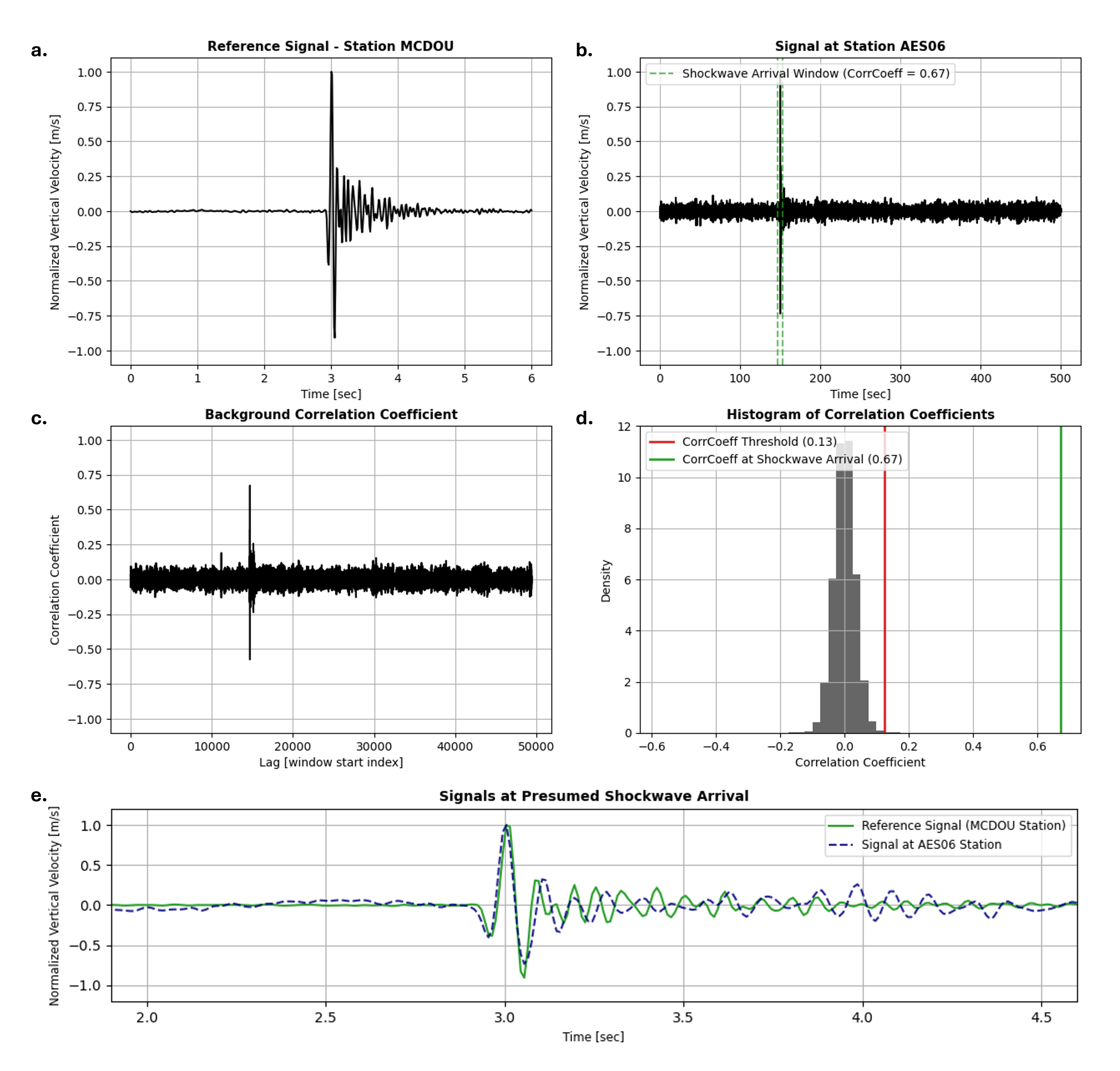}
    \caption{Hayabusa2 cross-correlation analysis. Seismic data are normalized, filtered between 1-19 Hz and represent the vertical ground velocity. \textbf{a.} Reference signal (6-second-long window of signal recorded at MCDOU (6K) station). \textbf{b.} Signal recorded at AES06 (5G) station, located 127.7 km from MCDOU station. The green dashed lines represent the containing the presumed shock wave arrival. \textbf{c.} Correlation coefficient as a function of the second signal window lag. \textbf{d.} Distribution of the correlation coefficients. The red line corresponds to the correlation coefficient threshold computed using the time-reversed template approach \citep{slinkard2014multistation} and with a false alarm rate of 0.1\%. \textbf{e.} Zoomed in within the window containing the presumed shock wave arrival. }
    \label{fig6}
\end{figure}

Cross-correlations were then performed between the 18 selected stations. Figure \ref{fig6} illustrates the cross-correlation between stations MCDOU (6K), used as the reference signal here, and station AES06 (5G). In figure \ref{fig6}b., the highlighted window corresponds to the presumed arrival of the shock wave in the AES06 signal, which we attribute to the Hayabsua2 re-entry. Additionally, Figure \ref{fig6}d. shows that the correlation coefficient for this window (0.67) is significantly higher than the threshold correlation coefficient (0.13). This suggests that if the reference signal originates from the Hayabusa2 SRC re-entry, AES06’s signal likely does as well.\\

\begin{figure}[h]
    \centering
    \includegraphics[width=0.98\textwidth]{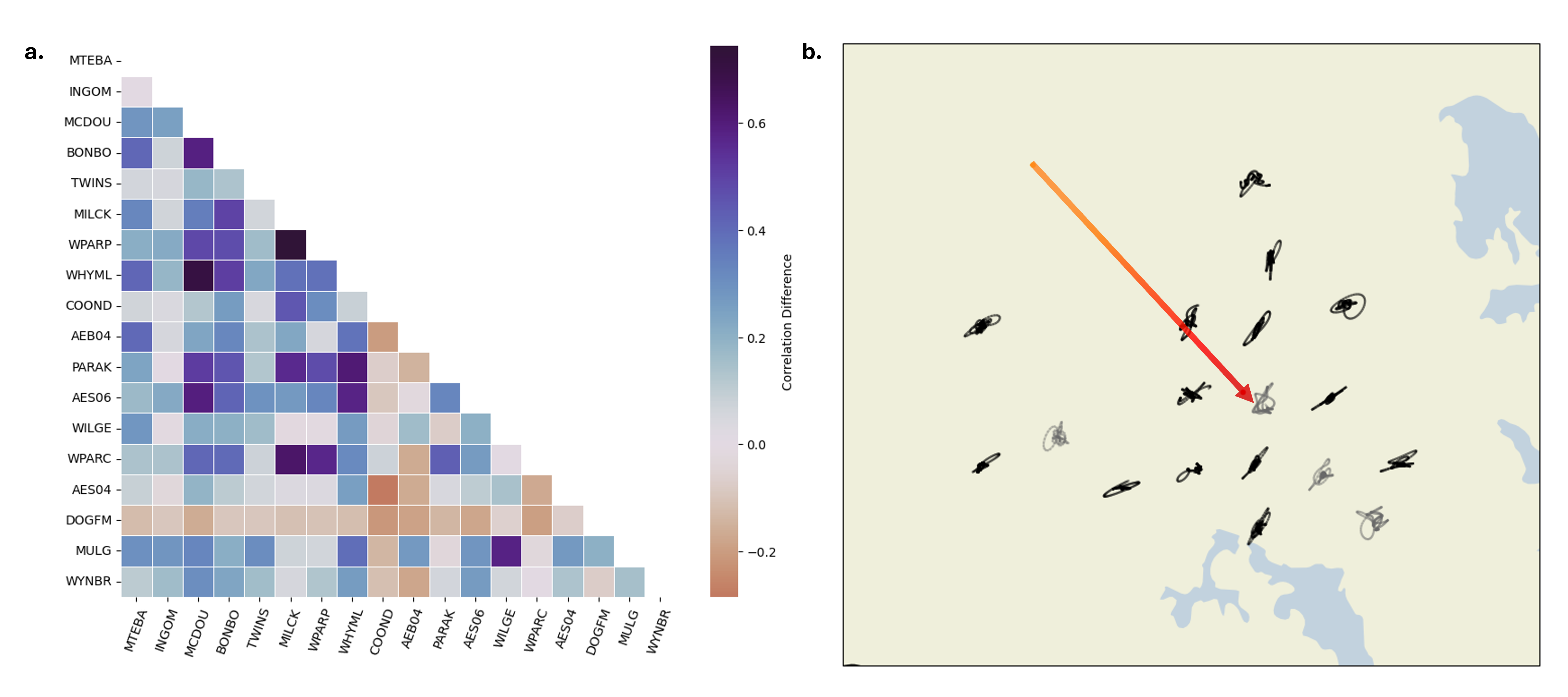}
    \caption{Results of the cross-correlation and polarization analyses of the Hayabusa2 SRC re-entry signal. \textbf{a.} Cross-correlation matrix showing the correlation coefficient after subtracting the background noise contribution. If the correlation difference value is below 0, the correlation may have been triggered by noise. \textbf{b.} Normalized particle motion of the Hayabusa2 re-entry signal recorded by 18 selected stations on the horizontal components (North vs East). The orange-red arrow indicates the SRC trajectory.}
    \label{fig7}
\end{figure}

\noindent Figure \ref{fig7}a. summarizes the results of the cross-correlation analysis for all 153 combinations of correlations performed between the 18 stations. It displays the difference between the threshold coefficient and the actual correlation coefficient for each station pair. Differences above 0 indicate that the signals from both stations are similar enough to be distinguished from the background noise. In total, 117 station pairs have cross-correlation coefficients above the threshold.\\
\noindent Figure \ref{fig7}a. also shows a decrease in correlations toward the lower right of the matrix, particularly beyond the WHYML station. DOGFM station also shows particularly poor results, with negative coefficients in its correlations with all other stations. As stations are sorted by their proximity to the re-entry trajectory (see also Table 2), this decrease in correlation coefficients is likely due to a distance limitation. The decrease after the WHYML station could therefore corresponds to a limit at approximately 80 kilometres. Other factors, such as local geology, which are not considered here, could also contribute.\\

Finally, in the polarisation analysis, particle motions of 14 of the 18 selected stations show notable horizontal orientations. The vertical component was excluded, as we have determined the signal originates from the atmosphere rather than any coupled or surface waves. Their motions point toward multiple points along the trajectory, which aligns with expectations for a ballistic shock wave origin, as presumed for Hayabusa2 SRC. For stations closer to the end of the re-entry, the particle motions appear to converge on a single point, which we interpret as the cone front. In contrast, stations higher along the re-entry path display orientations almost perpendicular to the trajectory (Figure \ref{fig7}b.). Of the four stations that show little preference, three are directly in line with the re-entry path.\\

As we see for this event, the dense distribution of sensors is fundamental to using this polarisation technique to distinguish likely source mechanisms. It cannot be generalised where seismic stations are typically sparse or inconsistently located with respect to sporadic events.\\

\noindent \textbf{Searching for common signals}
\vspace*{3mm} 

\begin{figure}[h]
    \centering
    \includegraphics[width=0.98\textwidth]{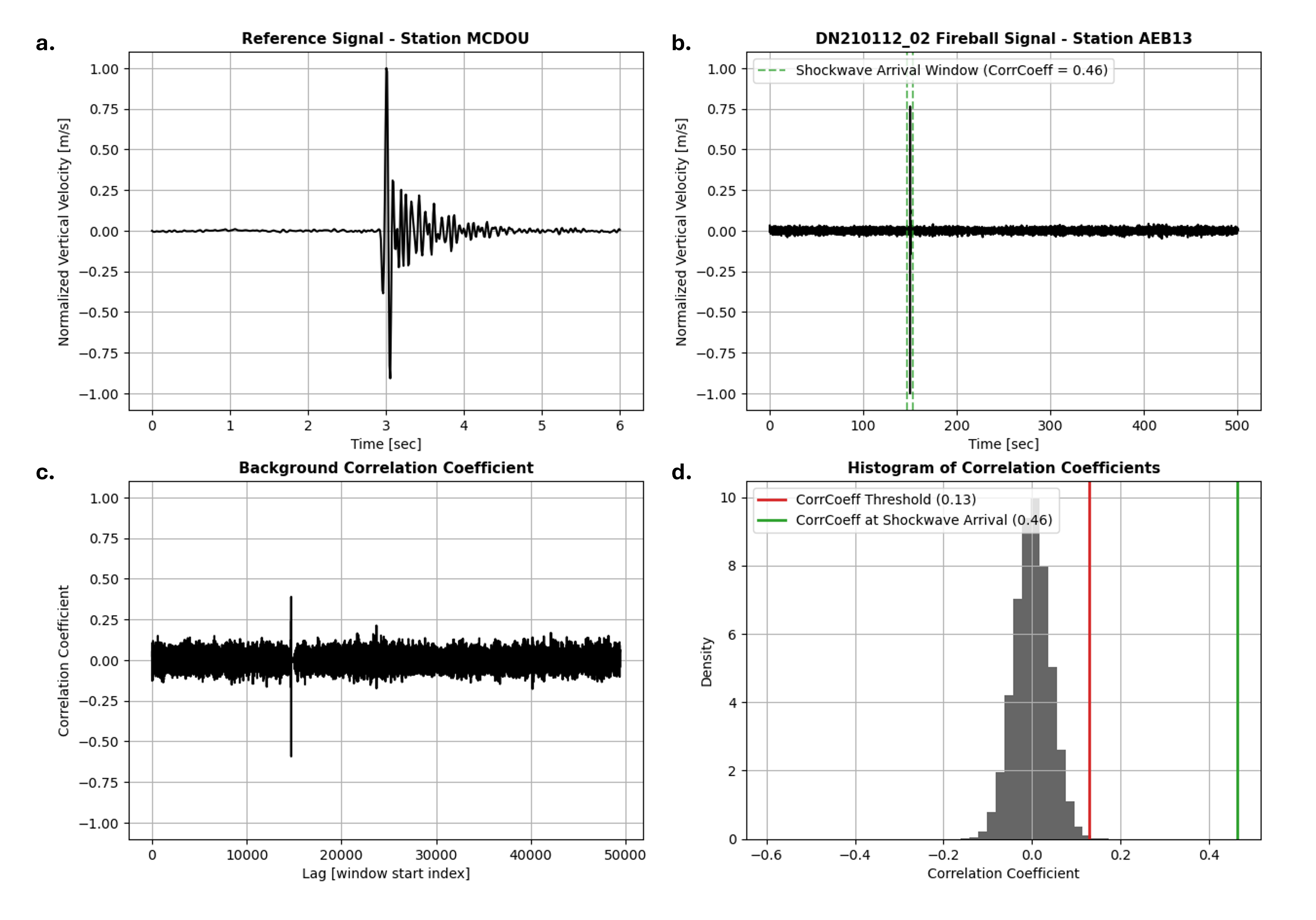}
    \caption{Cross-event signal correlation between the Hayabusa2 re-entry and DN210112\_02 fireball. Seismic data are normalized, filtered between 1-19 Hz and represent the vertical ground velocity. \textbf{a.} Reference signal (6-second-long window of signal recorded at MCDOU (6K) station). \textbf{b.} Signal from the DN210112\_02 fireball recorded at AEB13 (5G) station. The green dashed lines represent the containing the presumed shock wave arrival. \textbf{c.} Correlation coefficient as a function of the second signal window lag. \textbf{d.} Distribution of the correlation coefficients. The red line corresponds to the correlation coefficient threshold computed using the time-reversed template approach \citep{slinkard2014multistation} and with a false alarm rate of 0.1\%.}
    \label{fig8}
\end{figure}

\begin{figure}[h]
    \centering
    \includegraphics[width=0.98\textwidth]{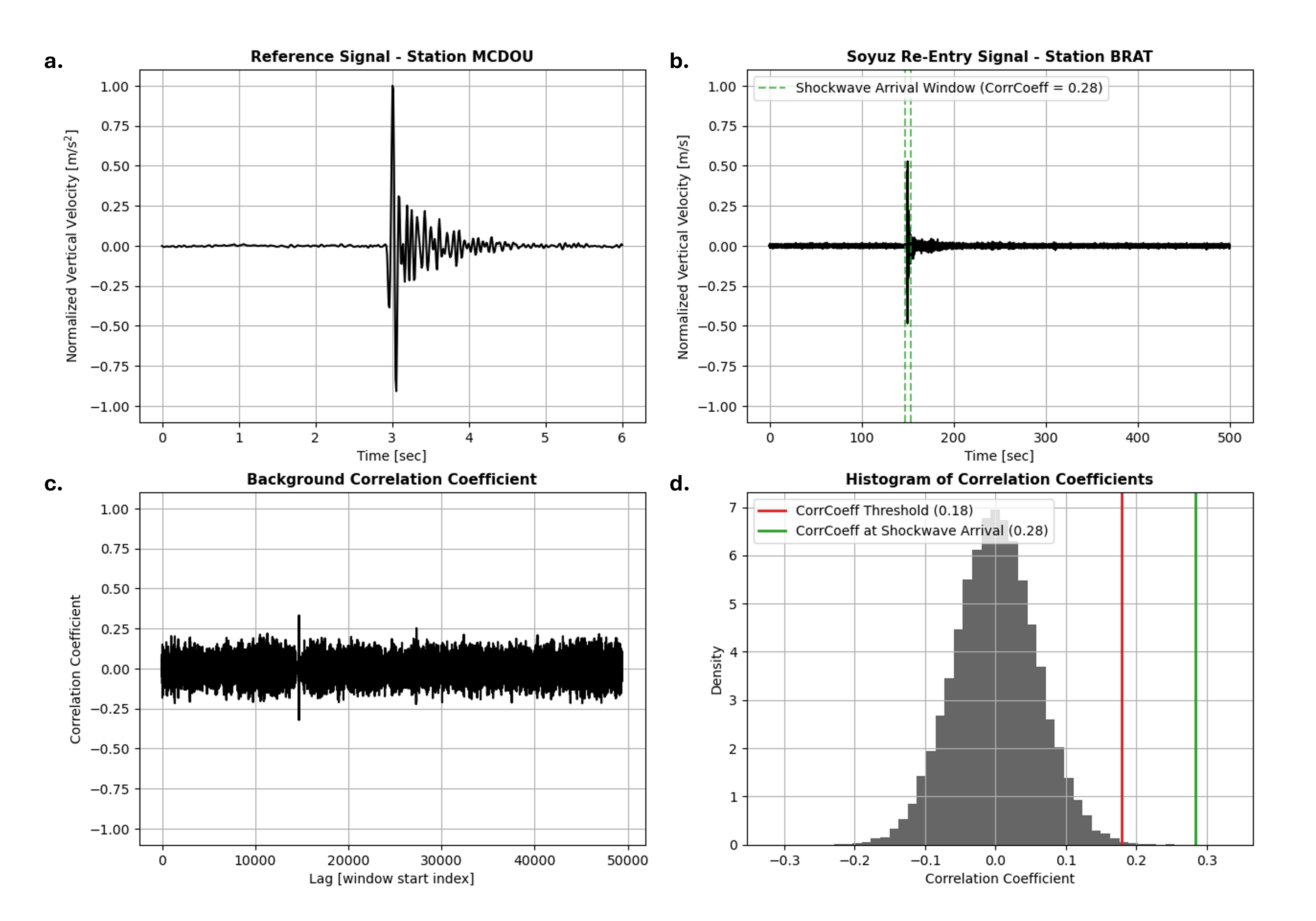}
    \caption{Cross-event signal correlation between the Hayabusa2 and Soyuz re-entries. Seismic data are normalized, filtered between 1-19 Hz and represent the vertical ground velocity. \textbf{a.} Reference signal (6-second-long window of signal recorded at MCDOU (6K) station). \textbf{b.} Signal from the Queensland fireball recorded at MTSU (AU) station. The green dashed lines represent the containing the presumed shock wave arrival. \textbf{c.} Correlation coefficient as a function of the second signal window lag. \textbf{d.} Distribution of the correlation coefficients. The red line corresponds to the correlation coefficient threshold computed using the time-reversed template approach \citep{slinkard2014multistation} and with a false alarm rate of 0.1\%.}
    \label{fig9}
\end{figure}

\begin{figure}[h]
    \centering
    \includegraphics[width=0.98\textwidth]{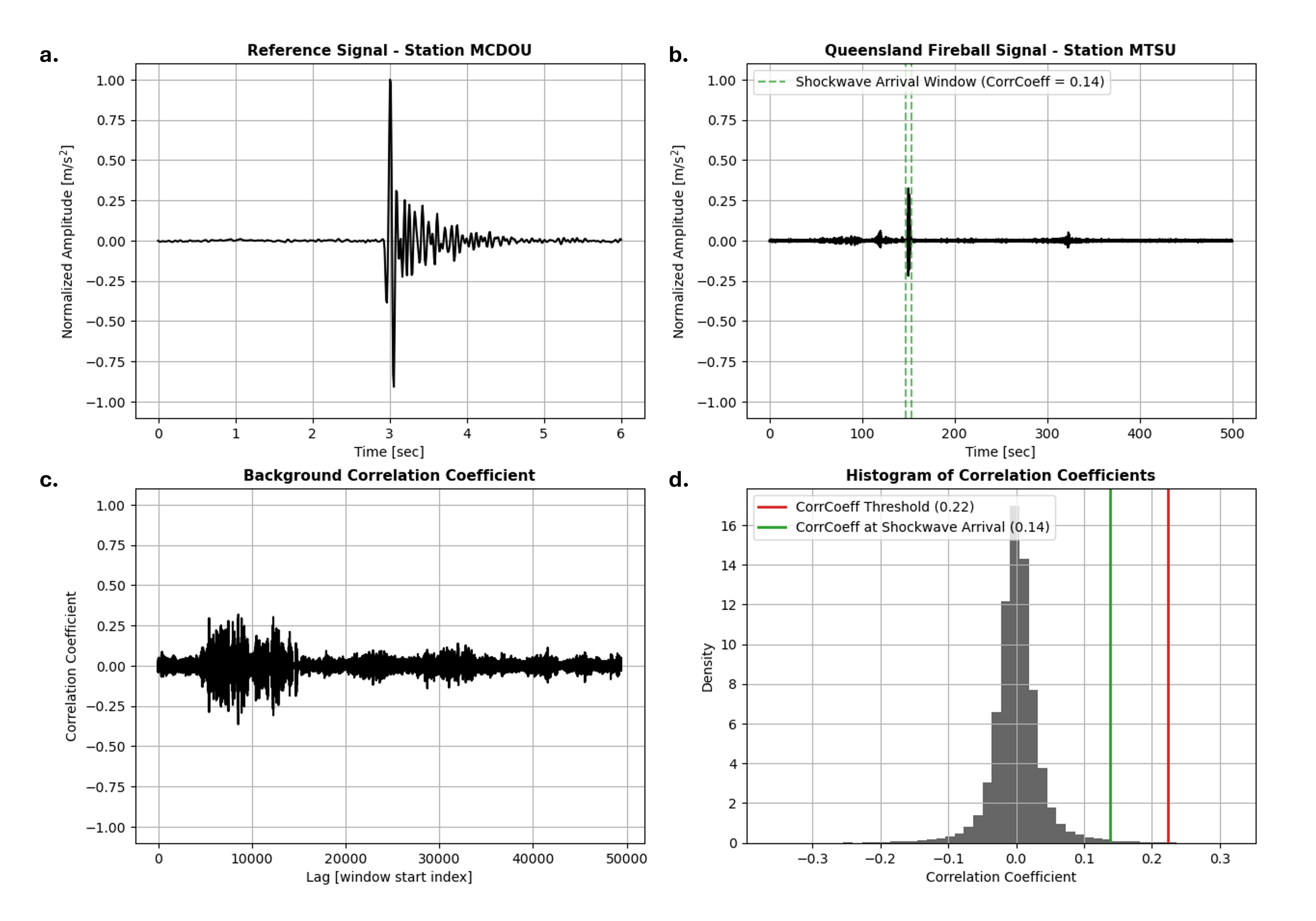}
    \caption{Cross-event signal correlation between the Hayabusa2 re-entry and the Queensland fireball. Seismic data are normalized, filtered between 1-19 Hz and represent the vertical ground velocity. \textbf{a.} Reference signal (6-second-long window of signal recorded at MCDOU (6K) station). \textbf{b.} Signal from the Soyuz re-entry recorded at BRAT (AU) station. The green dashed lines represent the containing the presumed shock wave arrival. \textbf{c.} Correlation coefficient as a function of the second signal window lag. \textbf{d.} Distribution of the correlation coefficients. The red line corresponds to the correlation coefficient threshold computed using the time-reversed template approach \citep{slinkard2014multistation} and with a false alarm rate of 0.1\%.}
    \label{fig10}
\end{figure}

The following Table 3 presents the results of comparing the Hayabusa2 SRC signal from the reference station MCDOU with those form other case study events: the DN210112\_02 fireball, the Soyuz re-entry and the Queensland fireball.\\
\noindent The cross-event signal correlation analysis yielded promising results when comparing Hayabusa2 SRC with both the DN210112\_02 fireball and the Soyuz re-entry. For DN210112\_02, the correlation coefficient threshold was 0.13, and the correlation with the window containing the presumed shock wave arrival returned a coefficient of 0.46, well above the false alarm rate (Figure \ref{fig8}). Similarly, for the Soyuz re-entry, a correlation coefficient of 0.28 was obtained in the window containing the presumed shock wave where there was a correlation coefficient threshold of 0.18 (Figure \ref{fig9}). In both cases, the correlation coefficients exceeded the threshold coefficients, indicating a significant similarity between the signals. In contrast, the cross-event signal correlation with the Queensland fireball, known to have had a significant airburst event, yielded a correlation coefficient of 0.14, below the threshold value of 0.22 (Figure \ref{fig10}).

\section*{Discussion and Conclusions}
\hspace{5mm} In this study, we began by analysing seismic records from the re-entry of the Hayabusa2 sample return capsule. To identify relevant signals, we employed a visualisation tool, based on the filterbanks technique, which splits the signal into narrow frequency bands. This approach allowed us to highlight the signal’s frequency content and helped determine the optimal frequency range for filtering the data. The connection between the observed signals and the SRC re-entry was then confirmed through a three-step validation process. This analysis demonstrated an average wave velocity falling within the speed of sound range, statistically significant correlation coefficients for most stations, and signal polarisations that align with the re-entry trajectory. The polarisation analysis also further reveals that the shock wave origin is the ballistic trajectory, more specifically from the Mach cone produced by the SRC’s hypersonic passage through the atmosphere. This finding is consistent with the dynamic characteristics known for the Hayabusa2 SRC; built to withstand atmospheric re-entry.\\

\noindent Given the confirmed ballistic shock wave origin (SM1), we used the Hayabusa2 signal as a reference in an attempt to identify other events with ballistic trajectory origins. Cross-correlations with signals from the other case studies events were therefore performed. Statistically significant correlation coefficients were obtained for both the DN210112\_02 fireball and the Soyuz re-entry using this cross-event signal correlation technique, suggesting that signals from these two events also originate from the ballistic trajectory. In contrast, the lower correlation observed for the Queensland fireball suggests a different origin. This aligns with the known dynamic characteristics of the Queensland event, which involved an airburst and likely fragmentation \citep{silber2024ground}. These phenomena generate spherically propagating shock waves that can interfere with each other as well as with the ballistic shock wave, potentially resulting in a more complex waveform. However, the analysis of this event is limited by the fact that only one station was available in the region of the Queensland bolide entry, restricting the depth of interpretations.\\
\\

Although factors such as the local geology, source-receiver distances, atmospheric conditions, impacting body and its trajectory characteristics or the fact that the waveform is a convolution of multiple sources were not considered in this study, we acknowledge their influence on signal waveforms and, consequently, on the cross-correlation results. Despite these unstudied effects, the technique of comparing the ballistic trajectory source of the Hayabusa2 SRC signal to other fireball events appears to yield promising results. High correlations were observed for the two events that lacked fragmentation/airburst (DN210112\_02 fireball and Soyuz re-entry), even though the Soyuz re-entry occurred in a different region that the Hayabusa2 re-entry. And as anticipated, a low correlation coefficient was found for the Queensland fireball, which an airburst.\\
\\

Once the dynamic behaviour of SRCs in the atmosphere is better understood and the technique refined to account for other factors such as the geological and atmospheric factors, this approach could provide valuable insights into dynamic entry processes for natural events occurring within the range of seismic networks, helping to determine whether a fireball underwent fragmentation/airburst or not. This is particularly useful in areas with limited optical network coverage. The method also shows promise for monitoring space debris re-entries. Space debris re-entries are a growing concern, and currently, there is no reliable way to verify whether a planned event has occurred over Australia aside from eyewitness reports or DFN observations. While planning debris re-entries, such as the Soyuz case study, have high time uncertainties, their ground track are generally well constrained. Hence, once improved, this method could be applied for template match searches within predicted re-entry windows. This capability would be particularly valuable for validating re-entries and refining re-entry time estimates. Another interesting future direction would be to construct event-specific filters by comparing the spectral content of the signal to the background noise, potentially improving the detection and isolation of SRC signatures.

\section*{Data and Resources}
Seismic data from the ‘AU’ network are publicly available via the FDSN service hosted by \citet{geoscience_australia} (https://auspass.edu.au). Seismic data from the ‘5G’ and ‘6K’ networks are restricted and not publicly available but can be accessed upon a request from their respective owners \citep{eakin2018lake, o2020ausarray}. The infraGA package is open-source and available at https://github.com/lanl-seismoacoustics/infraga (last accessed 14 February 2024; \citet{blom2012impulse, blom2017modeling}).

\section*{Declaration of competing interests}
The authors declare no competing interests.

\section*{Acknowledgments}
This work was funded by the Australian Research Council’s Discovery Project Scheme (DP230100301). We  would also like to thank Caroline Eakin and John Paul O’Donnell, the respective owners of the 5G and 6K seismic networks, for providing access to the data from their temporary networks and supporting our research, as well as the three reviewers of this manuscript for their helpful comments and suggestions.


\section*{Mailing addresses of authors}
\textbf{Iona Clemente; Eleanor K. Sansom; Hadrien A. R. Devillepoix:} Space Science and Technology Centre, School of Earth and Planetary Science, Curtin University, Perth, Australia \& International Centre for Radio Astronomy Research, Curtin University, Perth, Western Australia, Australia.\\
\noindent \textbf{Taichi Kawamura:} Institut de Physique du Globe de Paris, 35 Rue H\'el\`ene Brion, Paris 75013, France.\\
\noindent \textbf{Benjamin A. Fernando:} Department of Earth and Planetary Sciences, Johns Hopkins University, Baltimore, Maryland, United States.\\
\noindent \textbf{Raphael F. Garcia:} Institut Sup\'erieur de l’A\'eronautique et de l’Espace (ISAE-SUPAERO), Universit\'e de Toulouse, Toulouse, France.\\
\noindent \textbf{Olivia Collet:} Centre for Exploration Geophysics, Curtin University, GPO Box U1987, Perth, WA 6845, Australia.

\section*{Tables}
\small \textbf{Table 1:} Seismic stations and corresponding shortest distance ranges for all studied events.

\begin{center}
\footnotesize  
\begin{tabular}{l c c c c}
 \textbf{Events} & & \textbf{Number of Stations Used} & & \textbf{Distance Ranges (km)} \\ [0.4ex] 
 \hline\hline
 \textbf{Hayabusa2} (\textit{SRC re-entry}) & & 18$^{*}$ & & 39 - 184\\
 \hline
 \textbf{DN210112\_02 event} (\textit{Natural fireball}) & & 6$^{\ddag}$ & & 360 - 265\\
 \hline
 \textbf{Soyuz} (\textit{Spacecraft re-entry}) & & 4$^{\ddag}$ & & 65 - 127 \\
 \hline
 \textbf{Queensland fireball} (\textit{Natural fireball}) & & 1$^{\ddag}$ & & 198 \\
 \hline
 \footnotesize $^{*}$Cf. Table 2; 
 \footnotesize $^{\ddag}$Cf. Appendix A
\end{tabular}
\end{center}

\vspace*{15mm}

\noindent \small \textbf{Table 2:} Selected seismic stations for Hayabusa2 SRC re-entry analysis, with corresponding signal arrival times and shortest distances to the re-entry trajectory.

\begin{center}
\footnotesize  
\begin{tabular}{c c c c c}
 \textbf{Station Name (Network)} & & \textbf{Arrival Time (UTC hour)} & & \textbf{Distances to Trajectory (km)} \\ [0.4ex] 
 \hline\hline
 MTEBA (6K) & & 17:31:32 & & 39\\
 \hline
  INGOM (6K) & & 17:31:51 & & 50\\
 \hline
  MCDOU (6K) & & 17:32:07 & & 51\\
 \hline
  BONBO (6K) & & 17:32:10 & & 54\\
 \hline
  TWINS (6K) & & 17:32:28 & & 60\\
 \hline
  MILCK (6K) & & 17:32:34 & & 60\\
 \hline
  WPARP (6K) & & 17:32:04 & & 72\\
 \hline
  WHYML (6K) & & 17:33:34 & & 74\\
 \hline
  COOND (6K) & & 17:33:57 & & 95\\
 \hline
  AEB04 (5G) & & 17:34:16 & & 96\\
 \hline
  PARAK (6K) & & 17:33:58 & & 107\\
 \hline
  AES06 (5G) & & 17:34:58 & & 107\\
 \hline
  WILGE (6K) & & 17:35:55 & & 118\\
 \hline
  WPARC (6K) & & 17:34:16 & & 118\\
 \hline
  AES04 (5G) & & 17:35:38 & & 123\\
 \hline
  DOGFM (6K) & & 17:35:30 & & 123\\
 \hline
  MULG (AU) & & 17:36:09 & & 128\\
 \hline
  WYNBR (6K) & & 17:38:50 & & 184\\
 \hline
\end{tabular}
\end{center}
\vspace*{15mm}

\noindent \small \textbf{Table 3:} Cross-event signal correlation analysis outcomes.

\begin{center}
\footnotesize  
\begin{tabular}{l l c l c}
 \textbf{Events} & & \textbf{Airburst and/or Fragmentation} & & \textbf{Highest CC$^{*}$} \\ [0.4ex] 
 \hline\hline
 \textbf{DN210112\_02 event} (\textit{Natural fireball}) & & No & & 0.46\\
 \hline
 \textbf{Soyuz} (\textit{Spacecraft re-entry}) & & No & & 0.28 \\
 \hline
 \textbf{Queensland fireball} (\textit{Natural fireball}) & & Yes & & 0.14 \\
 \hline
 \footnotesize   $^{*}$CC: Correlation Coefficient
\end{tabular}
\end{center}

\section*{List of figures captions}
\small
\textbf{Figure 1:} Hayabusa2 sample return capsule re-entry trajectory.\\
\textbf{Figure 2:} Application of the filterbanks method on seismic records of the Hayabusa2 re-entry from the MCDOU (6K) station, a 3-component seismometer (Vertical, North, East).\\
\textbf{Figure 3:} Hayabusa2 SRC re-entry expected arrival time at MCDOU (6K) seismic station, with reference to the filterbanks filtering of the vertical signal component.\\
\textbf{Figure 4:} Hodochrone of seismic signals observed across the selected stations during Hayabusa2 SRC re-entry, plotted relatively to their shortest distances from the event (y-axis).\\
\textbf{Figure 5:} Outputs of infraGA model for ray tracing of shock waves, using the atmospheric model of \citet{nishikawa2022modeling}.\\
\textbf{Figure 6:} Hayabusa2 cross-correlation analysis.\\
\textbf{Figure 7:} Results of the cross-correlation and polarization analyses of the Hayabusa2 SRC re-entry signal.\\
\textbf{Figure 8:} Cross-event signal correlations between Hayabusa2 re-entry and DN210112\_02 fireball.\\
\textbf{Figure 9:} Cross-event signal correlations between Hayabusa2 and Soyuz re-entries.\\
\textbf{Figure 10:} Cross-event signal correlations between Hayabusa2 re-entry and the Queensland fireball.

\clearpage
\section*{Figures}
\captionsetup{font=small, labelfont=bf}

\captionsetup{font=small, labelfont=bf}

\captionsetup{font=small, labelfont=bf}

\captionsetup{font=small, labelfont=bf}

\captionsetup{font=small, labelfont=bf}

\captionsetup{font=small, labelfont=bf}

\captionsetup{font=small, labelfont=bf}

\captionsetup{font=small, labelfont=bf}

\captionsetup{font=small, labelfont=bf}

\captionsetup{font=small, labelfont=bf}

\clearpage
\section*{Appendix}
\textbf{Appendix A:} List of the seismic stations utilized for 3 of the case study events.

\begin{center}
\footnotesize  
\begin{tabular}{l l l}
 \textbf{Events} & & \textbf{Seismic Stations Used} \\ [0.4ex] 
 \hline\hline
 \textbf{DN210112\_02 event} (\textit{Natural fireball}) & & SOGAP, KOOTA, ARCOO, YUDNA, WHYML (6K), AEB13 (5G)\\
 \hline
 \textbf{Soyuz} (\textit{Spacecraft re-entry}) & & BRAT, GEXS, GVL, TOO (AU) \\
 \hline
 \textbf{Queensland fireball} (\textit{Natural fireball}) & & MTSU (AU) \\
 \hline
\end{tabular}
\end{center}

\end{document}